\documentclass[12pt]{article}
\usepackage{url}
\usepackage{amsfonts}
\usepackage{amsmath,amssymb,color}
\usepackage{graphicx}
\usepackage{enumitem}
\usepackage{multirow}
\usepackage{bm}
\usepackage{url}
\usepackage{amsfonts}
\usepackage{amsmath,amssymb,color}
\usepackage{graphicx}
\usepackage{enumitem}
\usepackage{multirow}
\usepackage{bm}
\usepackage{natbib}
\usepackage{verbatim}
\usepackage{float}
\usepackage[toc,page]{appendix}
\usepackage{setspace}
\usepackage{adjustbox}
\usepackage{rotating}
\usepackage{subcaption}
\usepackage{amsthm}
\usepackage{booktabs,dcolumn,caption}
\usepackage{fullpage}
\usepackage{array}

\newcolumntype{L}[1]{>{\raggedright\let\newline\\\arraybackslash\hspace{0pt}}p{#1}}
\newcolumntype{C}[1]{>{\centering\let\newline\\\arraybackslash\hspace{0pt}}p{#1}}
\newcolumntype{R}[1]{>{\raggedleft\let\newline\\\arraybackslash\hspace{0pt}}p{#1}}

\newcommand{\MC}{\mathbf B}
\newcommand{\MS}{\mathbf A}
\newcommand{\vc}{{\boldsymbol \beta}}
\newcommand{\vs}{{\boldsymbol \alpha}}

\newcommand{\bbb}{{\boldsymbol \beta}}

\newcommand{\SG}{\boldsymbol \Sigma}

\newcommand{\OOO}{\boldsymbol \Omega}

\newcommand{\BBB}{\mathcal B}

\newcommand{\bb}{\mbox{$\mathbf b$}}

\newcommand{\dd}{\mathbf d}

\newcommand{\EE}{\mathbf 1}

\newcommand{\JJ}{\mathbf J}
\newcommand{\MM}{\mathbf M}

\newcommand{\XX}{\mathbf X}

\newcommand{\NNN}{\mathcal N}

\newcommand{\II}{\mathbf I}

\newcommand{\LL}{\mathbf L}
\newcommand{\LLL}{\mathcal L}

\newcommand{\PPP}{\mathcal P}
\newcommand{\R}{\mathbb R}

\newcommand{\SSS}{\mathbf S}
\newcommand{\SSSS}{\mathcal S}

\newcommand{\TT}{\mathcal T}

\newcommand{\xx}{\mathbf x}

\newcommand{\1}{\uppercase\expandafter{\romannumeral1}}
\newcommand{\2}{\uppercase\expandafter{\romannumeral2}}
\newcommand{\rank}{\text{rank}}

\newcommand{\argmin}{\operatornamewithlimits{arg\,min}}

\newtheorem{theorem}{Theorem}

\newtheorem{algorithm}{Algorithm}

\newtheorem{lemma}{Lemma}

\begin{document}

\title{Signed Network Embedding with Application to Simultaneous Detection of Communities and Anomalies
}
\author{Haoran Zhang$^\dag$ and Junhui Wang$^\ddag$\\ [10pt]
	$^\dag$ Department of Statistics and Data Science \\
	Southern University of Science and Technology 
	\and
	$^\ddag$Department of Statistics \\
	The Chinese University of Hong Kong
}
\date{}
\maketitle

\onehalfspacing
\begin{abstract}
Signed networks are frequently observed in real life with additional sign information associated with each edge, yet such information has been largely ignored in existing network models. This paper develops a unified embedding model for signed networks to disentangle the intertwined balance structure and anomaly effect, which can greatly facilitate the downstream analysis, including community detection, anomaly detection, and network inference. The proposed model captures both balance structure and anomaly effect through a low rank plus sparse matrix decomposition, which are jointly estimated via a regularized formulation. Its theoretical guarantees are established in terms of asymptotic consistency and finite-sample probability bounds for network embedding, community detection and anomaly detection. The advantage of the proposed embedding model is also demonstrated through extensive numerical experiments on both synthetic networks and an international relation network.
\end{abstract}	


\doublespacing

\section{Introduction}\label{Sec:intro}

Network structure has been widely employed to describe pairwise interaction among a variety of objects. In literature, a number of popular models have been developed to leverage network structure for better modeling and prediction accuracy, including the Erd\"os-R\'enyi model \citep{Erdos1960evol}, the $\beta$-model \citep{Chatterjee2011rand, Graham2017econ}, stochastic block model \citep{holland1983stochastic, zhao2012consistency, sengupta2018block}, and network embedding model \citep{Hoff2002, zhang2021directed}.  Although success has been widely reported, most existing methods and theories are developed for unsigned network, and signed network has been largely ignored in literature.


One of the fundamental differences between signed network from unsigned network is the sign information associated with each edge, reflecting the polarity of node interaction. Examples of signed network include international politics \citep{heider1946attitudes, axelrod1993landscape, moore1979structural}, with consensus or conflict between different countries; and social networks \citep{massa2005controversial, leskovec2010predicting, kunegis2009slashdot}, with friends or enemies between users. The existence of negative edges leads to some unique structures in signed networks, such as the balance theory \citep{heider1946attitudes, cartwright1956structural}, which has made  most existing methods on unsigned network not directly applicable to signed network \citep{chiang2014prediction}. 

\begin{figure}[h]
	\centering
	\includegraphics[scale = 0.6]{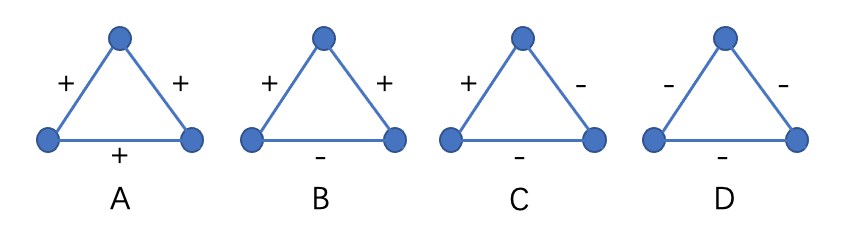}
	\caption{Balanced and unbalanced triads.}\label{fig:triad}
\end{figure}

In particular, the balance theory suggests that signed networks tend to conform to some local balanced patterns. A signed network is said to have strong balance if all its cycles have an even number of negative edges, following from the human intuition that  ``friend of friend is a friend" and ``enemy of enemy is also a friend" \citep{heider1946attitudes}. While strong balance implies clear-cut community structures \citep{harary1953notion}, it can be too restrictive and rarely satisfied in practice. Weak balance is proposed in \cite{davis1967clustering}, which only requires the signed network to have no cycle with exactly one negative edge and implies multiple communities in the signed network \citep{easley2010networks}. For illustration, Figure~\ref{fig:triad} displays four possible triads with signed edges, where triads A and C are strongly balanced, while triads A, C and D are weakly balanced. 


The balance theory provides additional guidance for community detection in a signed network. More specifically, besides sharing similar connectivity pattern, nodes in the same community tend to be connected with positive edges whereas nodes in different communities tend to be connected with negative edges \citep{tang2016survey}. In literature, a number of community detection methods for signed network have been developed \citep{doreian1996partitioning, bansal2004correlation,li2014comparative,chen2014overlapping, jiang2015stochastic,yang2007community}. However, most of them are algorithm oriented, and very little theoretical analysis has been conducted on the interplay between the balance theory and connectivity patterns.



Another distinctive feature of signed network is that violations of the balance theory are also prevalent \citep{cartwright1966number,bansal2004correlation,zheng2015social}. In other words, real-life signed networks can have triads like B in Figure \ref{fig:triad}, suggesting two friends of the same node can be enemies themselves.  For instance, Israel and Turkey are two close allies of the United States in international politics, but the relationship between themselves has not been very constructive. We refer to such violation as the anomaly effect, which is often encountered in real-life signed networks. It may convey important information that can not be explained by the balance theory, yet it has been largely ignored in the existing literature of signed network modeling. 


The major contributions of this paper are three-fold. First, to the best of our knowledge, this paper is one of the first attempts to incorporate both balance structure and anomaly effect in signed networks. Particularly, estimation of the balance structure can benefit substantially by taking the anomaly effect into account, and estimation of the anomaly effect per se is also of interest in various real applications, such as the international relation network in Section~\ref{sec:real}. Second, we propose a unified embedding model for signed networks to disentangle the intertwined balance structure and anomaly effect. It is cast into a flexible probabilistic model, where the balance structure and anomaly effect are modeled via a low rank plus sparse matrix decomposition. Finally, a thorough theoretical analysis is conducted to quantify the asymptotic estimation consistency of the proposed embedding model. We establish some novel identifiability conditions, under which both balance structure and anomaly effect can be consistently estimated with fast convergence rates. Its applications to community detection and anomaly detection in signed network are also considered, with sound theoretical justification. Particularly, under the signed stochastic block model (SSBM) with $n$ nodes, the proposed model achieves a fast convergence rate of $O_p(n^{-1})$ in terms of community detection, which matches up with the best existing results for unsigned network \citep{lei2015consistency}. In addition, the false discovery proportion of the proposed model also converges to 0 at a fast rate under some mild conditions.

The rest of paper is organized as follows. Section 2 presents the proposed embedding model for signed network as well as its estimation formulation and computational details. Section 3 establishes the theoretical results on the asymptotic consistencies of the proposed model, as well as the theoretical guarantees for its applications to community detection and anomaly detection. Section 4 conducts extensive numerical experiments on synthetic networks to examine the finite sample performance of the proposed model, and Section 5 applies it to analyze an international relation network. Section 6 concludes the paper, and technical proofs and necessary lemmas are provided in the Appendix.

Before moving to Section 2, we define some notations here. For a vector $\vc,$ let $\|\vc\|$ denote its Euclidean norm. For a matrix $\XX = (x_{ij}) = (\xx_1,...,\xx_n)^\top,$ we denote $\|\XX\|_0= \sum_{i,j}1_{\{x_{ij}\neq 0\}},~\|\XX\|_F= \sqrt{\sum_{i,j}x_{ij}^2},~\|\XX\|_{\max}= \max_{i,j}|x_{ij}|,$ and $\|\XX\|_{2\to\infty}= \max_{1\leq i\leq n}\|\xx_i\|.$ We also denote $\nu(\XX)$ and $\sigma_{k}(\XX)$ as the vectorization and the $k$-th largest singular value of $\XX,$ respectively.
Let $\II_n$ and $\EE_n$ denote the identity matrix of size $n$ and the vector with $n$ ones, respectively.


\section{Signed Network Embedding}

\subsection{Embedding Model}

Consider a signed network $\mathcal G$ with $n$ nodes labeled by $[n]=\{1,...,n\}$ and an adjacent matrix $\mathbf Y = (y_{ij})_{n\times n}$ with $y_{ij} \in \{-1,0,1\}$ and $y_{ij} = y_{ji}$. Here $y_{ij} = 1$ if there is a positive edge between node $i$ and node $j,$ $y_{ij} = -1$ if there is a negative edge, and $y_{ij} = 0$ if no edge is observed at all. Suppose the distribution function of $y_{ij}$ is given by
\begin{equation}
	\label{eq:model}
F(t\mid m_{ij}) := \Pr(y_{ij}\geq t\mid m_{ij}) =
\left\{\begin{aligned}
&1, ~~&t\leq -1,\\
&f(d_t + m_{ij}), ~~ &t\in \{0,1\},\\
&0, ~~ &t\geq 2,
\end{aligned}
\right.
\end{equation}
where $f$ is some pre-specified increasing link function such as the logit function or probit function, $\dd = (d_0,d_1)$ are intercepts, and $\MM = (m_{ij})_{n\times n}$ is an underlying matrix. Here a large value of $m_{ij}$ leads to a large probability of a positive edge between nodes $i$ and $j$, whereas a small value of $m_{ij}$ implies a large probability of a negative edge between nodes $i$ and $j$. It is also assumed that $y_{ij}$'s are mutually independent conditional on $\MM$. Furthermore, it is interesting to note that unsigned network can also be accommodated in \eqref{eq:model}, as the probability of negative edges becomes $0$ if $d_0$ is set as $\infty.$ 

To fully exploit the balance structure and anomaly effect in $\cal G$, we assume $\MM$ can be decomposed as $\MM = \LL + \SSS$, where $\LL = (l_{ij})_{n\times n}$ is a low rank matrix for the balance and community structure, and $\SSS = (s_{ij})_{n\times n}$ is a sparse matrix for the anomaly effect. Compared with the balance and community structures in $\LL$, it is believed that the level of the anomaly effect in $\SSS$ is much weaker, and it is only observed on a relatively small number of edges \citep{facchetti2011computing}, leading to the sparsity in $\SSS$. 


The modeling strategy for $\LL$ is motivated from the fact that both weak balance and community structure can be naturally accommodated via network embedding in a low dimensional space.  Specifically, we set $l_{ij} = - \|\vc_i-\vc_j\|^2$ with $\vc_i \in\R^{K_1}$ being the embedding vector for node $i$, which makes $\LL$ a negative Euclidean distance matrix with $\rank(\LL)\leq K_1+2.$ We refer readers to Chapter 5 of \cite{dattorro2010convex} for the rank of a Euclidean distance matrix. With the embedding vectors, two nodes with positive edge tend to have a small distance in the embedding space, whereas two nodes with negative edge have a relatively large distance. As a direct consequence, triads A, C and D in Figure~\ref{fig:triad} are allowed under this embedding framework, whereas triad B is forbidden due to the triangle inequality. Besides weak balance, this Euclidean embedding framework also encourages nodes with similar connectivity patterns to be situated in a close neighborhood in the embedding space. In the sequel, we denote the corresponding parameter space for $\LL$ as \begin{equation}\label{eq:community space}
\LLL_n = \{\LL = (l_{ij})\in\R^{n\times n}: l_{ij} = - \|\vc_i-\vc_j\|^2,~\MC = (\vc_1,...,\vc_n)^\top\in\R^{n\times K_1}\}.
\end{equation}


Modeling of $\SSS$ requires additional structural assumption, since it is generally difficult to construct consistent estimate of $\SSS$ from a single observed $\cal G$ with random noise \citep{chandrasekaran2011rank, candes2011robust}. To see this, given that $\LL$ is known in prior, we only have one  observation $y_{ij}\in\{-1,0,1\}$ to estimate each $s_{ij}$, making it almost impossible unless further structure on $\SSS$ is imposed. Particularly, we assume that $\SSS\in\SSSS_n$ with 
\begin{equation}
\label{eq:anomaly space}
\SSSS_n = \{\SSS\in\R^{n\times n}: \SSS=\MS\MS^\top,~\MS\in\R^{n\times K_2}\},
\end{equation} 
where $\MS = (\vs_1,...,\vs_n)^\top,$ and $s_{ij} = \vs_i^\top \vs_j$ with $\vs_i$ being an additional embedding vector of node $i,$ determining whether it may have anomalous edges with other nodes.



\subsection{Estimation Formulation}\label{subsec:est}

Let $\TT = \{-1,0,1\}$ denote the support of $y_{ij}$. For each $t\in \TT,$ we further denote the probability of $y_{ij} = t$ as 
$$
p(t \mid m_{ij}) := 
F(t\mid m_{ij}) - F( t+1\mid m_{ij}). 
$$ 
Then the log likelihood of the signed network $\mathcal G$ takes the form 
\begin{equation}
\begin{aligned}
\log L (\MM) &= \sum_{i,j=1}^n\log p(y_{ij} \mid m_{ij}) 
&=\sum_{i,j=1}^n \log\left[ F(y_{ij}\mid m_{ij}) - F(y_{ij}+1\mid m_{ij}) \right]. \nonumber
\end{aligned}
\end{equation}

Given the embedding framework in \eqref{eq:community space} and \eqref{eq:anomaly space}, we rewrite $\log L(\MM)=\log L(\MC,\MS)$ with $m_{ij} = - \|\vc_i-\vc_j\|^2 + \vs_i^\top\vs_j$, and propose the estimation formulation as
\begin{equation}\label{eq:opti}
\begin{aligned}
(\widehat\MC,\widehat\MS) = &\argmin_{\MC\in\R^{n\times K_1},\MS\in\R^{n\times K_2}} \left\{- \log L(\MC,\MS) \right\} \\
\text{s.t.}~~  &\|\MC\|_{2\to\infty}  \leq C,~~\|\MS\|_{2\to\infty}  \leq C,
~~ \EE_n^\top\MC = \textbf{0},~~ \EE_n^\top\MS = \textbf{0},\\
&\MC^\top\MS = \textbf{0},~~\text{and}~~\|\MS\|_F \leq \kappa\sqrt{a_n}\|\MC\|_F.
\end{aligned}
\end{equation} 
Here, $C$ and $\kappa$ are some pre-specified constants, and $a_n$ is a small anomaly rate that controls the relative scales of $\|\MC\|_F$ and $\|\MS\|_F,$ which reflects the prior knowledge that the level of anomaly effect is much weaker than that of the balance structure. The sum-to-zero and orthogonal constraints on $\MC$ and $\MS$ are necessary for their identifiability in terms of parameter estimation. 
Given $(\widehat\MC,\widehat\MS),$ we define $\widehat\LL = (\widehat l_{ij})_{n\times n}$ with $\widehat l_{ij} = - \|\widehat\vc_i - \widehat\vc_j\|^2,$
$\widehat\SSS = \widehat\MS\widehat\MS^\top$ and $\widehat\MM = \widehat\LL + \widehat\SSS.$ 


Once $\widehat\MC$ and $\widehat\SSS$ are obtained, we can further detect communities and anomalies in the signed network. Particularly, we perform an $(1+\epsilon)$-approximation of the $K$-means algorithm \citep{kumar2004simple} on the estimated $\{\widehat\vc_i\}_{i=1}^n$ to detect communities, and then $$
\widehat\NNN_l = \{i\in[n]:\widehat\psi_i = l\}, ~~\text{for}~~  l=1,...,m,
$$ is the $l$th detected community, where $m$ denotes the number of communities, and $\widehat \psi_i \in[m]$ denotes the community membership of node $i.$  We also detect anomalies by performing hard thresholding on $\widehat s_{ij},$ and conclude the edge $y_{ij}$ to be anomalous if $|\widehat s_{ij}| > \eta_n,$ where $\eta_n$ is the thresholding parameter.

\subsection{Computation}\label{subsec:compute}
 
The optimization task in \eqref{eq:opti} can be efficiently solved by an alternative updating scheme, which updates $\MC$ and $\MS$ iteratively via the projected gradient descent algorithm.  Specifically, for a matrix $\XX$ and positive constant $c,$ we define some projection operators as following $$
\PPP_{F,c}(\XX) := 1_{\{\|\XX\|_{F}\leq c\}}\XX + 1_{\{\|\XX\|_{F}>c\}}c\|\XX\|_{F}^{-1}\XX,
$$ $$
\PPP_{2\to\infty,c}(\XX) := 1_{\{\|\XX\|_{2\to\infty}\leq c\}}\XX + 1_{\{\|\XX\|_{2\to\infty}>c\}}c\|\XX\|_{2\to\infty}^{-1}\XX.
$$ 
Further, given $\MC\in\R^{n\times K_1},$ we define $$
\PPP_{\MC}^\perp(\MS) := \left( \II_n - \MC(\MC^\top\MC)^{-1}\MC^\top \right)\MS,
$$ 
for any $\MS\in\R^{n\times K_2},$ which is the projection operator onto the orthogonal complement of $\MC$'s column space.


Then given $\left(\MC^{(k)},\MS^{(k)}\right)$ with $\EE_n^\top\MS^{(k)} = \bf0,$ and $\{d_0,~d_1,~C,~\kappa,~a_n\},$  we implement the following updating scheme: 
\begin{equation}\label{eq:update BA}
\begin{aligned}
&\MC^{(k+1)} = \PPP_{2\to\infty,C}\left[\JJ_n\left(\MC^{(k)} + \xi_1\frac{\partial \log L(\MC^{(k)},\MS^{(k)})}{\partial\MC}\right)\right],\\
&\MS^{(k+1)} = \PPP_{2\to\infty,C}\left\{\PPP_{F,\kappa\sqrt{a_n}\|\MC^{(k+1)}\|_F}\left[\PPP^\perp_{(\MC^{(k+1)},\EE_n)}\left(\MS^{(k)} + \xi_2\frac{\partial \log L(\MC^{(k+1)},\MS^{(k)})}{\partial\MS}\right)\right]\right\},
\end{aligned}
\end{equation}
where $\xi_1,\xi_2>0$ are step sizes and $\JJ_n = \II_n - \EE_n\EE_n^\top/n$.  We repeat the above updating steps until convergence to get $(\widehat\MC,\widehat\MS).$

If the intercepts $\dd$ are unknown a priori, their estimates can be obtained in the iterative updating algorithm as well. We denote $\log L(\MC,\MS,\dd)$ to emphasize the dependence of the log likelihood on $\dd,$ and define $\PPP_{[a,b]}(x) = x 1_{\{a\leq x\leq b\}} + a1_{\{x<a\}} + b1_{\{x>b\}},$ for any interval $[a,b]$ and scalar $x.$ Given $\dd^{(k)}=(d_0^{(k)},d_1^{(k)})$ satisfying $c_1\leq d_1^{(k)}\leq d_0^{(k)}+\delta<d_0^{(k)}\leq c_2,$ where $\delta>0,c_1,c_2$ are pre-specified constants, in addition to the updating steps in \eqref{eq:update BA}, we also update $\dd$ as
\begin{equation*}
\begin{aligned}
& d_1^{(k+1)} = \PPP_{[c_1,d_0^{(k)}-\delta]}\left(d_1^{(k)} + \xi_3 \frac{\partial \log L(\MC^{(k+1)},\MS^{(k+1)},\dd^{(k)})}{\partial d_1}\right),\\
& d_0^{(k+1)} = \PPP_{[d_1^{(k+1)}+\delta,c_2]}\left(d_0^{(k)} + \xi_4 \frac{\partial \log L(\MC^{(k+1)},\MS^{(k+1)},(d_0^{(k)},d_1^{(k+1)}))}{\partial d_0}\right),
\end{aligned}
\end{equation*}
where $\xi_3,\xi_4>0$ are step sizes.

It is clear that the embedding performance of \eqref{eq:opti} relies on the choices of $K_1$ and $K_2,$ which can be determined by some data-adaptive tuning procedure, such as the network cross-validation \citep{li2020network}. If community detection is of primary interest, we would suggest to set $K_1= K_2 = m-1,$ where only the number of communities $m$ is determined by some tuning procedure \citep{saldana2017many}. Furthermore, we could also set $C=2$ and $\kappa=1$, while the value of $a_n$ can be set by the practitioners to reflect their prior knowledge about the scale of anomaly effect in the given signed network. 


\section{Theory}\label{sec:theory}

Denote $(\LL^*,\SSS^*)\in\LLL_n\times\SSSS_n$ as the true parameters, and $\MM^* = \LL^*+\SSS^*.$ Further denote $\MC^*\in\R^{n\times K_1}$ and $\MS^*\in\R^{n\times K_2}$ as the embedding vectors for $\LL^*$ and $\SSS^*,$ respectively.  For any constant $\alpha > 0$, we define 
$$
G_{\alpha} := \max_{t\in\TT}\sup_{|x|\leq\alpha} \frac{\left| p'(t\mid x) \right|}{p(t\mid x)}
~~\text{and}~~H_{\alpha} := \min_{t\in\TT}\inf_{|x|\leq\alpha} \left\{ \frac{[ p'(t\mid x) ]^2}{[p(t\mid x)]^2} - \frac{ p''(t\mid x) }{p(t\mid x)} \right\},
$$ 
where $p'(t\mid x)$ and $p''(t\mid x)$ denote, respectively, the first and second order derivatives of $p(t\mid x)$ with respect to $x$. 

Recall the constants $C,\kappa$ and sequence $a_n\to0$ in \eqref{eq:opti}, and we make the following technical conditions.


\textbf{Condition A1.} $\|\MC^*\|_{2\to\infty}\leq C,~\|\MS^*\|_{2\to\infty} \leq C$ and $G_{5C^2+d_0} < \infty,~H_{5C^2+d_0} > 0$.

\textbf{Condition A2.} $\EE_n^\top\MC^*= \textbf{0},~\EE_n^\top\MS^* = \textbf{0},$ and $(\MC^*)^\top\MS^* = \textbf{0}.$

\textbf{Condition A3.} There exist two positive definite matrices $\SG_1$ and $\SG_2$ such that $\text{tr}(\SG_2) < \kappa^2\text{tr}(\SG_1)$ and 
$$
\begin{aligned}
\frac{1}{n}(\MC^*)^\top\MC^* \to \SG_1,&&~~\text{with}~~\lambda_{1}(\SG_1) > ... > \lambda_{K_1}(\SG_1) > 0,\\
\frac{1}{na_n}(\MS^*)^\top\MS^* \to \SG_2,&&~~\text{with}~~\lambda_{1}(\SG_2) > ... > \lambda_{K_2}(\SG_2) > 0.
\end{aligned}
$$


Condition A1 assumes that the true embedding vectors are all constrained in a compact set, and the probability function $p(t\mid \cdot)$ is neither too steep nor too flat in the feasible domain, which is satisfied by most common link functions, such as the logit and probit functions. Similar conditions have also been assumed in \cite{bhaskar2016probabilistic} for quantized matrix completion. Condition A2 assumes that the embedding vectors are centralized, and the two embedding matrices $\MC^*$ and $\MS^*$ are orthogonal to each other.  Condition A3 quantifies the difference in the relative scales of the true balance structure and the anomaly effect. It holds true with high probability if $\vc_i^*$ are independent copies from a mean zero distribution in $\R^{K_1},$ whose covariance matrix $\SG_1$ has $K_1$ different positive eigenvalues; and $\vs_i^*$ are independent copies from a mixture distribution $(1-a_n)f_0 + a_nf_1,$ where $f_0$ is the Dirac delta distribution with point mass at $\bf0,$ and $f_1$ is a mean zero distribution in $\R^{K_2}$ whose covariance matrix $\SG_2$ has $K_2$ different positive eigenvalues.

Lemma~\ref{lem:ident} establishes the identifiability of the balance structure $\LL$ and anomaly effect $\SSS$.

\begin{lemma}\label{lem:ident}
Suppose Conditions A2 and A3 hold. Further suppose there exist $\widetilde\MC\in\R^{n\times K_1}$ and $\widetilde\MS\in\R^{n\times K_2}$ such that $\EE_n^\top\widetilde\MC= \textbf{0},~\EE_n^\top\widetilde\MS = \textbf{0},~\widetilde\MC^\top\widetilde\MS = \textbf{0},~\widetilde\MC^\top\widetilde\MC/n \to \SG_1$ and $\widetilde\MS^\top\widetilde\MS/(na_n) \to \SG_2$. If $-\|\widetilde\vc_i-\widetilde\vc_j\|^2 + \widetilde\vs_i^\top\widetilde\vs_j = -\|\vc_i^*-\vc_j^*\|^2 + (\vs_i^*)^\top\vs_j^*$ for all $(i,j)\in[n]\times[n]$, then for $n$ large enough, it holds that $\|\widetilde\vc_i-\widetilde\vc_j\| = \|\vc_i^*-\vc_j^*\|$ and $\widetilde\vs_i^\top\widetilde\vs_j = (\vs_i^*)^\top\vs_j^*$ for all $(i,j)$.
\end{lemma}


Now, we establish the upper bounds on the estimation errors of $\widehat\MM$ and $(\widehat\LL,\widehat\SSS)$.


\begin{theorem}\label{thm:M rate}
Suppose Conditions A1-A3 hold. Then, with probability at least $1-C_1\exp(-C_2n)$, it holds true that \begin{equation}\label{eq:M rate}
\frac{1}{n}\|\widehat\MM - \MM^*\|_F \leq r_n,
\end{equation} 
where $r_n = 4.02\sqrt{2(K+2)n^{-1}}G_{5C^2+d_0}H_{5C^2+d_0}^{-1},$ $C_1$ and $C_2$ are universal constants, and $K = K_1+K_2.$ 
\end{theorem}



\begin{theorem}\label{thm:LS rate}
Suppose Conditions A1-A3 hold. Then, there exists a constant $c$ such that, with probability at least $1-C_1\exp(-C_2n),$ 
\begin{equation}\label{eq:L rate}
\frac{1}{n}\|\widehat\LL - \LL^*\|_F \leq c\frac{C\sqrt{2K_1(K+2)}G_{5C^2+d_0}H_{5C^2+d_0}^{-1}}{\sqrt{n}},
\end{equation} 
\begin{equation}\label{eq:S rate}
\frac{1}{n}\|\widehat\SSS - \SSS^*\|_F \leq c\frac{C\sqrt{2K_2(K+2)}G_{5C^2+d_0}H_{5C^2+d_0}^{-1}}{\sqrt{na_n}}.
\end{equation} 
\end{theorem}


Theorems \ref{thm:M rate} and \ref{thm:LS rate} shows that both $\MM^*$ and $(\LL^*,\SSS^*)$ can be consistently estimated by $\widehat\MM$ and $(\widehat\LL,\widehat\SSS),$ respectively. Note that the convergence rate of $\widehat\SSS-\SSS^*$ depends on the anomaly rate $a_n,$ which reflects the scale of anomaly effect.

We then turn to establish the asymptotic consistency in terms of community detection. The following condition on the true community structure is required.


\textbf{Condition A4.}
There exists a finite set $\BBB = \{\bb_1,...,\bb_m\} \subset \R^{K_1}$ such that $\bbb_i^* \in \BBB,$ for any $i=1,...,n.$ 


Condition A4 is equivalent to the signed stochastic block model \citep{jiang2015stochastic}, which assumes that the embedding vector of each node is fully determined by its community membership.
Let $\psi^*_i\in[m]$ be the community index for node $i$ such that $\bbb_i^* = \bb_{\psi^*_i}.$ Then the $l$-th community is defined as 
$$
\NNN_l^* = \{i\in[n]:\psi_i^* = l\}, ~~ \text{for} ~~ l=1,...,m,
$$ 
which is invariant up to some permutation of the community index. 

Further, define $\xi_n = \min_{l\in[m]} |\NNN_l^*|/n$ as the proportion of nodes in the smallest community, and $\zeta_n = \min_{1\leq l<k\leq m} \|\bb_l-\bb_k\|$ as the minimal distance between $\bb_l$'s. Both $\xi_n$ and $\zeta_n$ are allowed to converge to 0 as $n$ diverges.



\begin{theorem}\label{thm:community}
Suppose Conditions A1-A4 hold. Further suppose \begin{equation}\label{eq:comm cond}
K_1r_n^2 = o(\xi_n\zeta_n^2),
\end{equation} 
where $r_n$ is defined as in Theorem~\ref{thm:M rate}. 
Then, there exists a constant $c$ such that, with probability at least $1-C_1\exp(-C_2n),$ \begin{equation}\label{eq:comm error1}
\min_{\{p_1,...,p_m\}}\frac{1}{n} \sum_{l=1}^m |\NNN_l^*\setminus \widehat\NNN_{p_l}| \leq c\frac{K_1(K+2)G_{5C^2+d_0}^2H_{5C^2+d_0}^{-2}}{n\zeta_n^2},
\end{equation} 
and \begin{equation}\label{eq:comm error2}
\min_{\{p_1,...,p_m\}}\max_{1\leq l\leq m} \frac{|\NNN_l^*\setminus \widehat\NNN_{p_l}|}{|\NNN_l^*|} \leq c\frac{K_1(K+2)G_{5C^2+d_0}^2H_{5C^2+d_0}^{-2}}{n\xi_n\zeta_n^2},
\end{equation}
where $\{p_1,...,p_m\}$ is a permutation of $[m]$. 
\end{theorem}

In Theorem \ref{thm:community}, \eqref{eq:comm error1} gives a bound for overall proportion of mis-clustered nodes, while \eqref{eq:comm error2} gives a bound for the worst case proportion of mis-clustered nodes in each communities. To guarantee detection consistency, \eqref{eq:comm cond} requires that the quantity $\xi_n\zeta_n^2$ converges to 0 at a speed not faster than $K_1r_n^2.$  If $\xi_n=\Omega(1),~\zeta_n = \Omega(1),$ and $K_1,K_2,C$ are fixed, then the convergence rates in both \eqref{eq:comm error1} and \eqref{eq:comm error2} are of order $n^{-1},$ which matches up with the best existing results for unsigned network in literature \citep{lei2015consistency}. 



Finally, we establish the asymptotic consistency in terms of anomaly detection. The following condition on the sparsity of $\SSS^*$ is required.

\textbf{Condition A5.}
There exists a universal constant $c_0>0$ such that 
\begin{equation}\label{eq:sparsity}
c_0^{-1}n^2a_n^2\leq\|\SSS^*\|_0 \leq c_0n^2a_n^2,
\end{equation}
\begin{equation}\label{eq:min}
 C\sqrt{K_2}r_na_n^{-\frac{3}{2}} = o( s_{\min}),
\end{equation}
where $r_n$ is defined as in Theorem~\ref{thm:M rate},  and $s_{\min} = \min_{s_{ij}^*\neq0}|s_{ij}^*|.$


In Condition A5, \eqref{eq:sparsity} assures the number of nonzero entries in $\SSS^*$ is of order $n^2a_n^2,$ and \eqref{eq:min} requires that the minimal absolute value of the nonzero entries in $\SSS^*$ is not too close to zero. Then, with a proper choice of $\eta_n,$ Theorem~\ref{thm:anomaly} establishes an upper bound for the false discovery proportion of $\widehat\SSS.$



\begin{theorem}\label{thm:anomaly}
Suppose Conditions A1-A3 and A5 hold and the thresholding parameter $\eta_n$ is set so that 
\begin{equation}\label{eq:thresholding}
\eta_n = o(s_{\min})~~\text{and}~~C\sqrt{K_2}r_na_n^{-\frac{3}{2}} = o(\eta_n),
\end{equation}
where $r_n$ is defined as in Theorem~\ref{thm:M rate}. Then, there exists a constant $c$ such that, with probability at least $1-C_1\exp(-C_2n),$ 
\begin{equation}\label{eq:FDP rate}
\frac{\#\{(i,j):|\widehat s_{ij}| > \eta_n,~s_{ij}^*=0\}}{\#\{(i,j):|\widehat s_{ij}| > \eta_n\} \vee 1} \leq c\frac{C^2K_2(K+2)G_{5C^2+d_0}^2H_{5C^2+d_0}^{-2}}{na_n^3\eta_n^2}.
\end{equation} 
\end{theorem}


It is clear that the upper bound in \eqref{eq:FDP rate} converges to 0 as $n$ diverges, 
whose convergence rate is governed by $n,a_n$ and $\eta_n.$ Particularly, when $K$ and $C$ are fixed, the false discovery proportion of $\widehat\SSS$ converges to zero at a fast rate of $n^{-1/2}\log n,$ provided that $a_n^3\eta_n^2$ is set as the same order of $(\sqrt{n}\log n)^{-1}.$

\section{Numerical Experiments}\label{sec:simu}

We examine the finite-sample performance of the proposed method in terms of both community detection and anomaly detection. For community detection, we compare the proposed method, denoted as SNE, with two existing signed network community detection methods in literature \citep{chiang2012scalable,cucuringu2019sponge}, as well as a naive embedding method which only considers the balance structure $\LL$ while ignoring the anomaly effect $\SSS$; denoted as BNC, SPONGE and naive, respectively. Their community detection accuracy is measured by the overall community detection error rate in \eqref{eq:comm error1}. Furthermore, as existing methods rarely consider anomaly detection, we just report the false discovery proportion \eqref{eq:FDP rate} of SNE in different scenarios to demonstrate its effectiveness in anomaly detection.

The following two synthetic networks are considered.

{\bf Example 1.} The synthetic network is generated from the SSBM model with anomalies. Specifically, we set $m=4$, and generate $\psi_i^*$ from a multinomial distribution on $\{1,2,3,4\}$ with probability $(0.1,0.2,0.3,0.4).$ Let $\vc_{i}^* = \bb_{\psi_i^*} \sim N(\textbf{0},\II_{3})$, and $\vs_{i}^*\sim (1-a_n)f_0 + a_nf_1,$ where $f_0$ is the Dirac delta distribution with point mass at $\bf0$ and $f_1 = 0.5N(\EE_3, \OOO_2)+0.5N(-\EE_3, \OOO_2)$ is a mixture Gaussian distribution with $\OOO_2$ to be a $3\times 3$ diagonal matrix with diagonal entries independently generated from a uniform distribution on $[0,0.1].$


{\bf Example 2.} The synthetic network is generated from a mixture model with anomalies. Specifically, for community structure, we set $m=4$, and generate $\psi_i^*$ from a multinomial distribution on $\{1,2,3,4\}$ with probability $(0.25,0.25,0.25,0.25).$ Let $\vc_{i}^* \sim N(\bb_{\psi_i^*},0.01\II_3)$ with $\bb_{\psi_i^*} \sim N(\textbf{0},\II_{3})$, and $\vs_{i}^*$ is generated similarly as in Example 1. 


Various scenarios in each example are considered, with $a_n \in \{0, 0.1,0.2,0.3\}$, and $n \in \{200,500,1000\}$. For each scenario, the averaged community detection errors for all the methods over 50 independent replications, together with their standard errors, are reported in Tables \ref{tab:comm1} and \ref{tab:comm2}. 

\begin{table}[!h]
\begin{center}
\caption{The averaged community detection errors of various methods over 50 independent replications and their standard errors  in Example 1.}
\label{tab:comm1}
\begin{small}
\begin{tabular}{ c|c|c|c|c|c } 
\hline
\hline
 & Method & $a_n = 0$ & $a_n = 0.1$ & $a_n = 0.2$ & $a_n = 0.3$  \\
\hline
\hline
\multirow{4}{4em}{$n=200$} & SNE &\textbf{0.0426}(0.0054) &0.0507(0.0051)& \textbf{0.0914}(0.0076)& \textbf{0.0787}(0.0067)  \\ \cline{2-6} 
& naive & \textbf{0.0426}(0.0054) & \textbf{0.0443}(0.0043) & 0.1051(0.0079) & 0.1649(0.0043)  \\ \cline{2-6} 
& BNC & 0.1333(0.0034) & 0.1302(0.0042) & 0.1253(0.0027) & 0.1348(0.0038) \\ \cline{2-6} 
& SPO & 0.2213(0.0032) & 0.2143(0.0034) & 0.1381(0.0066) & 0.1926(0.0032) \\ \cline{2-6} 
\hline
\hline
\multirow{4}{4em}{$n=500$} & SNE & \textbf{0.0062}(0.0018) & \textbf{0.0120}(0.0025) & \textbf{0.0074}(0.0018) & \textbf{0.0844}(0.0042)  \\ \cline{2-6} 
& naive & \textbf{0.0062}(0.0018) & 0.0174(0.0031) & 0.0462(0.0028) & 0.1518(0.0024) \\ \cline{2-6} 
& BNC & 0.1162(0.0013) & 0.1282(0.0026) & 0.1207(0.0014) & 0.1263(0.0018) \\ \cline{2-6} 
& SPO & 0.2252(0.0015) & 0.1010(0.0030) & 0.1834(0.0027) & 0.1984(0.0012) \\ \cline{2-6} 
\hline
\hline
\multirow{4}{4em}{$n=1000$} & SNE & \textbf{0.0058}(0.0013) & \textbf{0.0115}(0.0018) & \textbf{0.0083}(0.0013) & \textbf{0.0640}(0.0024) \\ \cline{2-6} 
& naive &\textbf{0.0058}(0.0013) & 0.0173(0.0022) & 0.0461(0.0014) & 0.1619(0.0013)\\ \cline{2-6} 
& BNC &0.1232(0.0017) & 0.1099(0.0008) & 0.1148(0.0011) & 0.1279(0.0012) \\ \cline{2-6} 
& SPO &0.2144(0.0020) & 0.1461(0.0007) & 0.1942(0.0010) & 0.1546(0.0017) \\ \cline{2-6} 
\hline
\hline
\end{tabular}
\end{small}
\end{center}

\end{table}

\begin{table}[!h]
\begin{center}
\caption{The average community detection errors of various methods over 50 independent replications and their standard errors  in Example 2.}
\label{tab:comm2}
\begin{small}
\begin{tabular}{ c|c|c|c|c|c } 
\hline
\hline
 & Method & $a_n = 0$ & $a_n = 0.1$ & $a_n = 0.2$ & $a_n = 0.3$  \\
\hline
\hline
\multirow{4}{4em}{$n=200$} & SNE &\textbf{0.0326}(0.0047)& 0.0601(0.009)& 0.0496(0.0067)& \textbf{0.0795}(0.0068)  \\ \cline{2-6} 
& naive &\textbf{0.0326}(0.0047)& \textbf{0.0519}(0.008)& \textbf{0.0463}(0.0066)& 0.0965(0.0045)  \\ \cline{2-6} 
& BNC &0.2991(0.0056)& 0.3260(0.007)& 0.2923(0.0051)& 0.3023(0.0049) \\ \cline{2-6} 
& SPO &0.2031(0.0113)& 0.1694(0.012)& 0.0710(0.0086)& 0.1251(0.0092) \\ \cline{2-6} 
\hline
\hline
\multirow{4}{4em}{$n=500$} & SNE &\textbf{0.0330}(0.0048)& 0.0359(0.0047)& \textbf{0.0227}(0.0028)& \textbf{0.0648}(0.0034)  \\ \cline{2-6} 
& naive &\textbf{0.0330}(0.0048)& \textbf{0.0249}(0.0027)& 0.0484(0.0036)& 0.1606(0.0043)  \\ \cline{2-6} 
& BNC &0.3274(0.0057)& 0.3160(0.0056)& 0.2438(0.0032)& 0.2623(0.0035) \\ \cline{2-6} 
& SPO &0.0828(0.0065)& 0.0306(0.0026)& 0.1425(0.0057)& 0.2283(0.0052) \\ \cline{2-6} 
\hline
\hline
\multirow{4}{4em}{$n=1000$} & SNE &\textbf{0.0191}(0.0029)& 0.0308(0.0035)& \textbf{0.0294} (0.0028)& \textbf{0.0777} (0.0045)  \\ \cline{2-6} 
& naive &\textbf{0.0191}(0.0029)& \textbf{0.0279}(0.0019)& 0.0387 (0.0035)& 0.0808 (0.0041)  \\ \cline{2-6} 
& BNC &0.3282(0.0040)& 0.2921(0.0035)& 0.2496 (0.0053)& 0.2628 (0.0058) \\ \cline{2-6} 
& SPO &0.0315(0.0035)& 0.0726(0.0037)& 0.1047 (0.0049)& 0.0969 (0.0047) \\ \cline{2-6} 
\hline
\hline
\end{tabular}
\end{small}
\end{center}

\end{table}


It is clear from Tables \ref{tab:comm1} and \ref{tab:comm2} that SNE outperforms the other three competitors in most scenarios. Particularly, SNE and naive yield the same numerical performance when $a_n=0$, but the advantage of SNE becomes more and more substantial when both $n$ and $a_n$ increase, which confirms the benefit of incorporating the anomaly effects into the signed network modeling. Furthermore, both BNC and SPO do not produce satisfactory and stable numerical performance, in that SPO yields reasonable performance in Example 2, but much worse performance than other methods in Example 1. 

In addition, the averaged false discovery proportions of SNE over 50 independent replications and its standard error are reported in Table \ref{tab:anom}. It is evident that its false discovery proportion decreases as $n$ grows, confirming the asymptotic convergence estabilished in Theorem~\ref{thm:anomaly}. It is also interesting to remark that the performance of SNE in anomaly detection deteriorates as $a_n$ increases, which is possibly due to the fact that the difference between the anomaly effect and balance structure shrinks as $a_n$ grows, making anomaly detection more challenging.

\begin{table}[!htb]
	\begin{center}
		\caption{The averaged false discovery proportions of SNE over 50 independent replications and its standard errors.}
		\label{tab:anom}
		\begin{tabular}{ c|c|c|c|c } 
			\hline
			\hline
			&  & $a_n = 0.1$ & $a_n = 0.2$ & $a_n = 0.3$  \\
			\hline
			\hline
			\multirow{3}{5em}{Example 1} & $n=200$ &0.9736(0.0017) & 0.7253(0.0081) & 0.4320(0.0143)  \\ \cline{2-5} 
			& $n=500$ &0.2349(0.0038) & 0.1538(0.0081) & 0.2117(0.0059)  \\ \cline{2-5} 
			& $n=1000$ &0.0022(0.0003) & 0.0585(0.0043) & 0.1831(0.0052) \\ \cline{2-5} 
			\hline
			\hline
			\multirow{3}{5em}{Example 2} & $n=200$ &0.9651(0.0030)& 0.7859(0.0125)& 0.6496(0.0216) \\ \cline{2-5} 
			& $n=500$ &0.2788(0.0063)& 0.1103(0.0043)& 0.2365(0.0099) \\ \cline{2-5} 
			& $n=1000$ &0.0287(0.0034)& 0.0290(0.0018)& 0.1910(0.0057) \\ \cline{2-5} 
			\hline
			\hline
		\end{tabular}
	\end{center}
\end{table}

\section{International Relation Network}\label{sec:real}

We now apply the proposed SNE method to analyze the international relation network, which is constructed based on the Correlates of War dataset during 1993—2014 \citep{maoz2019dyadic}. In this network, we set $y_{ij} = -1$ if there was ever a conflict between countries $i$ and $j$, $y_{ij} = 1$ if countries $i$ and $j$ were always in alliance and never had conflict during the whole time period, and $y_{ij}=0$ if countries $i$ and $j$ had neither alliance nor conflict. This leads to a signed network with 152 nodes, 2026 positive edges  and 694 negative edges. 

We first set $K_1 = K_2 = m-1$, and employ the Bayesian information criterion \citep{saldana2017many} to select $m=6$, which is consistent with some existing studies \citep{traag2009community, jiang2015stochastic}. We then apply SNE with $C=2,~\kappa = 1,~a_n = 0.1$ on this signed network to obtain the embedding vectors $\widehat\MC$ and $\widehat\SSS$. We further perform an $(1+\epsilon)$-approximation of the K-means algorithm on $\{\widehat\vc_i\}_{i=1}^n$, and obtained the estimated community membership $\widehat\NNN_l$ for $l=1,\ldots,6$. In addition, we set $\eta_n$ as the median of the absolute value of all $\widehat s_{ij}$'s, and denote $\widetilde s_{ij} = \widehat s_{ij} 1_{\{|\widehat s_{ij}|>\eta_n\}}.$ 

\begin{figure}[!h]
  \centering
    \begin{subfigure}[b]{0.45\textwidth}
        \centering
        \includegraphics[height=2.5in]{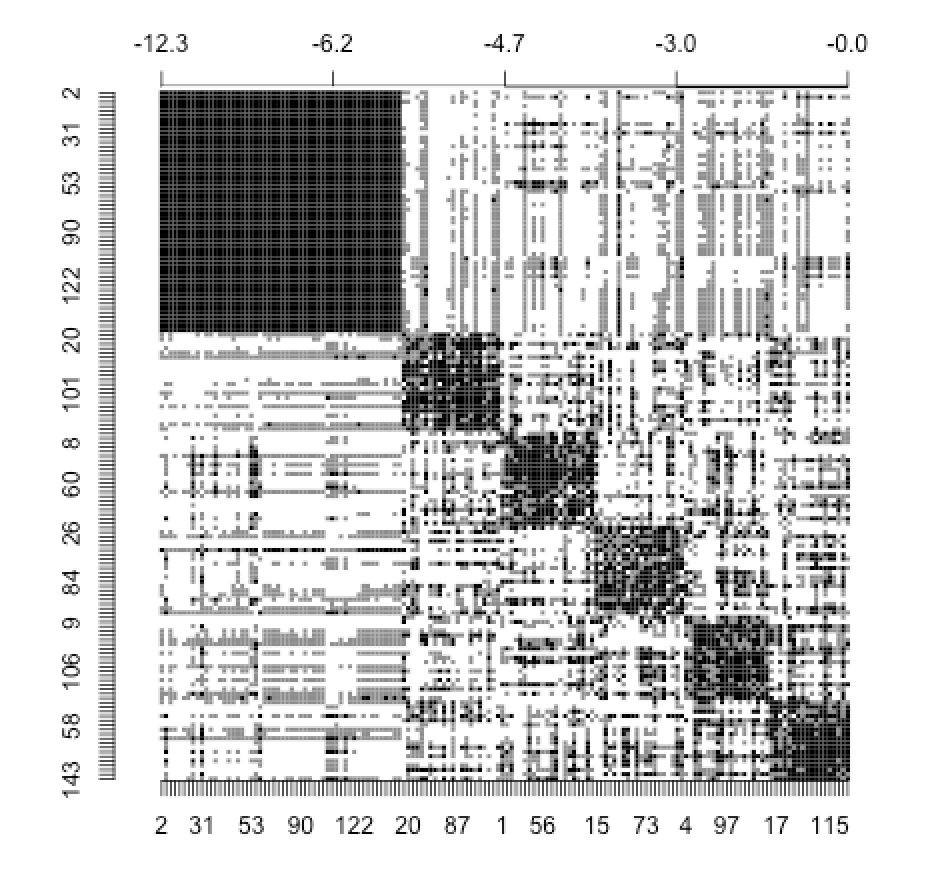}
    \end{subfigure}
    \begin{subfigure}[b]{0.45\textwidth}
        \centering
        \includegraphics[height=2.5in]{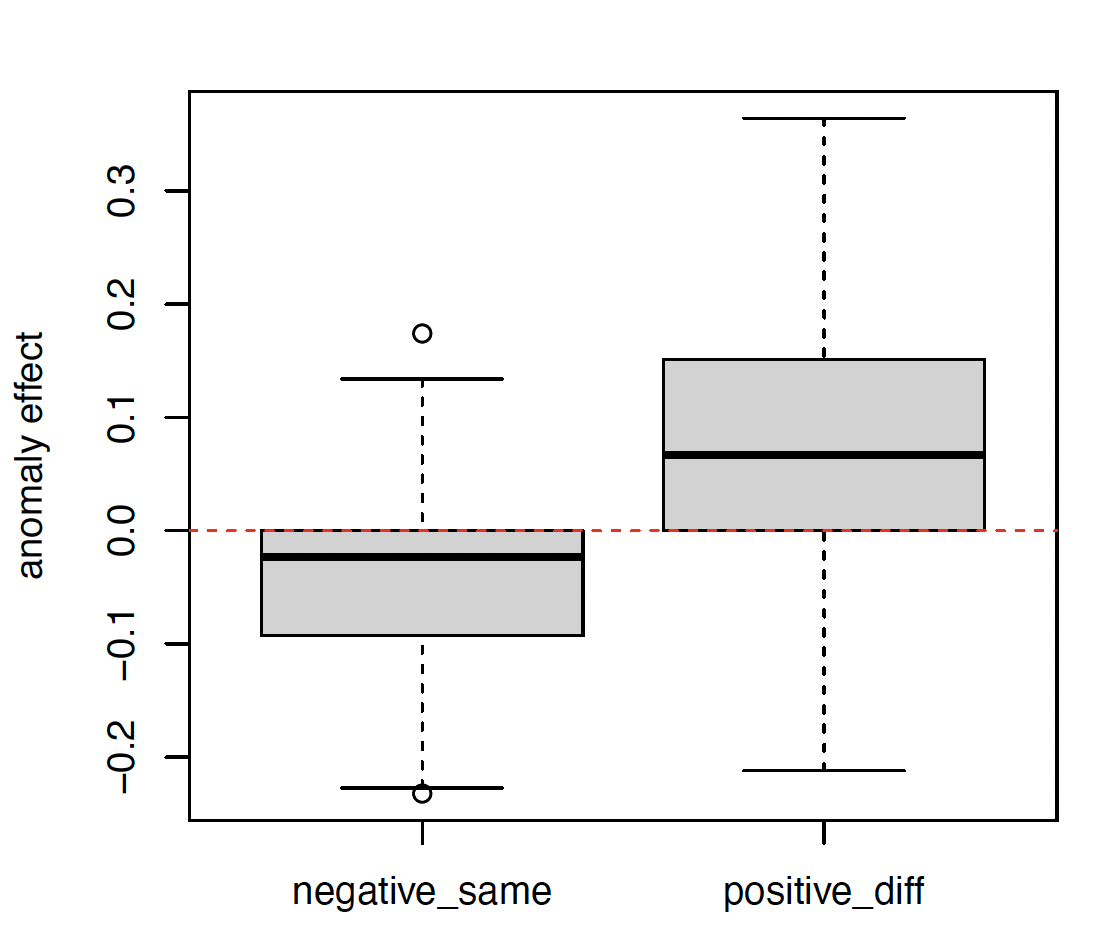}
    \end{subfigure}
 \caption{The heatmap for the rearranged $\widehat\LL$ according to the estimated community membership is displayed in Panel (a), and a side-by-side boxplot for $\widetilde s_{ij}$ is displayed in Panel (b).}
 \label{fig:results}
\end{figure}

Displayed in Panel (a) of Figure~\ref{fig:results} is the heatmap of a rearranged $\widehat\LL$ according to $\widehat\NNN_l$, showing a clear block diagonal structure produced by SNE. Here the darker the color is, the larger $\widehat l_{ij}$ is, which also indicates a smaller distance $\|\widehat\vc_i - \widehat\vc_j\|$. Displayed in  Panel (b) of Figure~\ref{fig:results} is a side-by-side boxplot of $\widetilde s_{ij}$, where the left boxplot shows the distribution of $\widetilde s_{ij}$ with $i,j$ in the same community but $y_{ij} = -1$, and the right boxplot shows the distribution of $\widetilde s_{ij}$ with $i,j$ in different communities but $y_{ij} = 1$. 
It is evident that most $\widetilde s_{ij}$'s in the left boxplot are negative, while those $\widetilde s_{ij}$'s in the right boxplot tend to be positive, confirming the validity of SNE in anomaly detection.


\begin{figure}[!h]
	\centering
	\includegraphics[scale = 1.1]{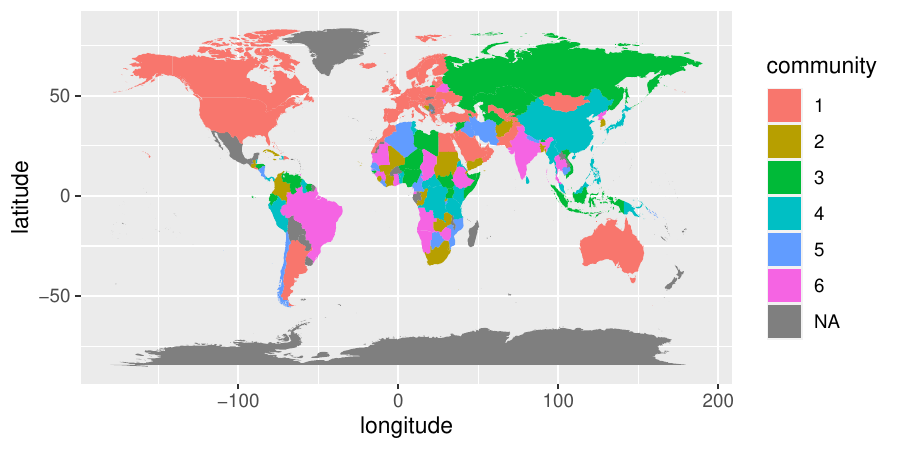}
	\caption{World map with countries in different communities.}
	\label{fig:map}
\end{figure}

We also color the world map according to the estimated community membership in  Figure \ref{fig:map}, where countries colored in grey are not included in the dataset. The detailed country list of each community can be found in Appendix C. It is clear from Figure \ref{fig:map} that the first and largest community contains the United States and its political allies, including most western European countries, Canada, Australia and some countries in Middle East such as Israel, Saudi Arabia and United Arab Emirates.  The third community consists of Russia and its political allies including Yugoslavia, Vietnam and some countries in central Asia such as Turkmenistan and Kyrgyzstan. The fourth community consists of countries in East Asia including China and Japan, and countries in central Africa including Tanzania and Central African Republic. It is interesting to note the community structure in Africa is rather complicated, which is probably due to the frequent conflicts between African countries during the time period.



\begin{figure}[!h]
  \centering
    \begin{subfigure}[b]{0.45\textwidth}
        \centering
	\includegraphics[scale = 0.2]{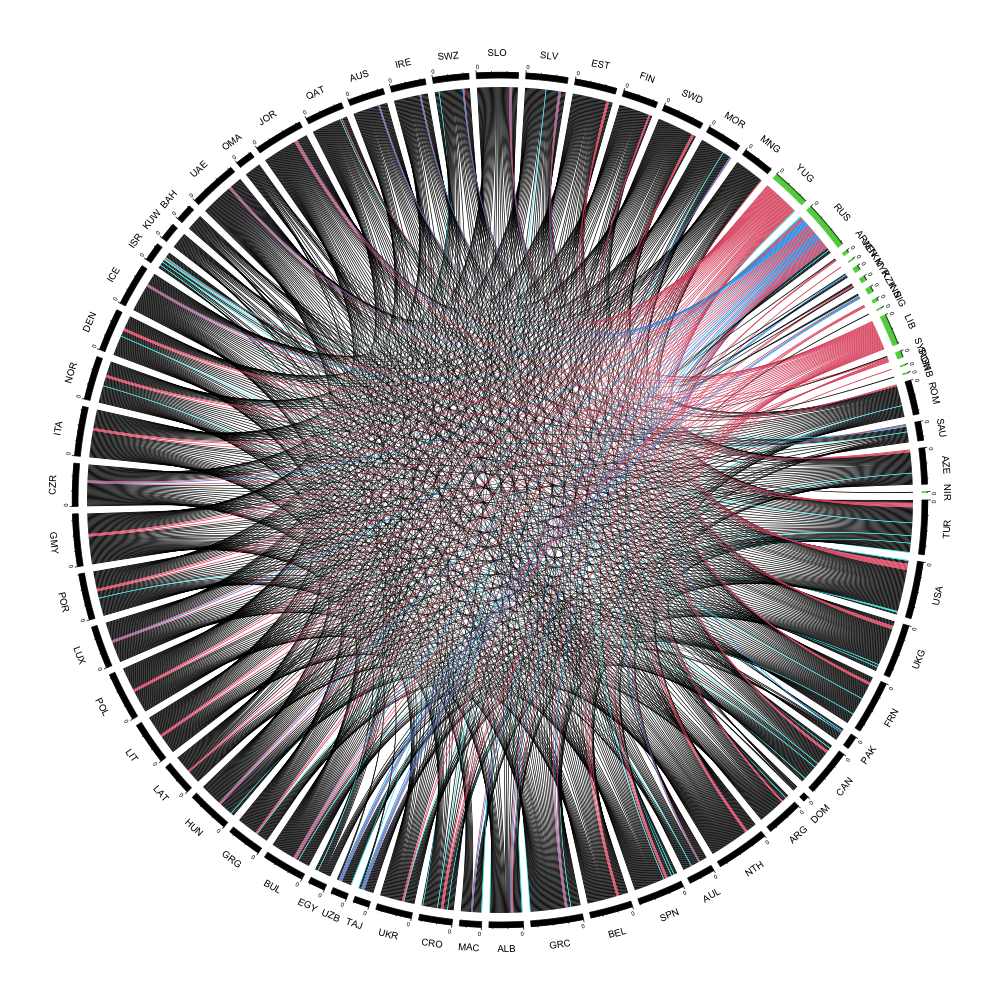}
    \end{subfigure}
    \begin{subfigure}[b]{0.45\textwidth}
        \centering
        \includegraphics[scale = 0.2]{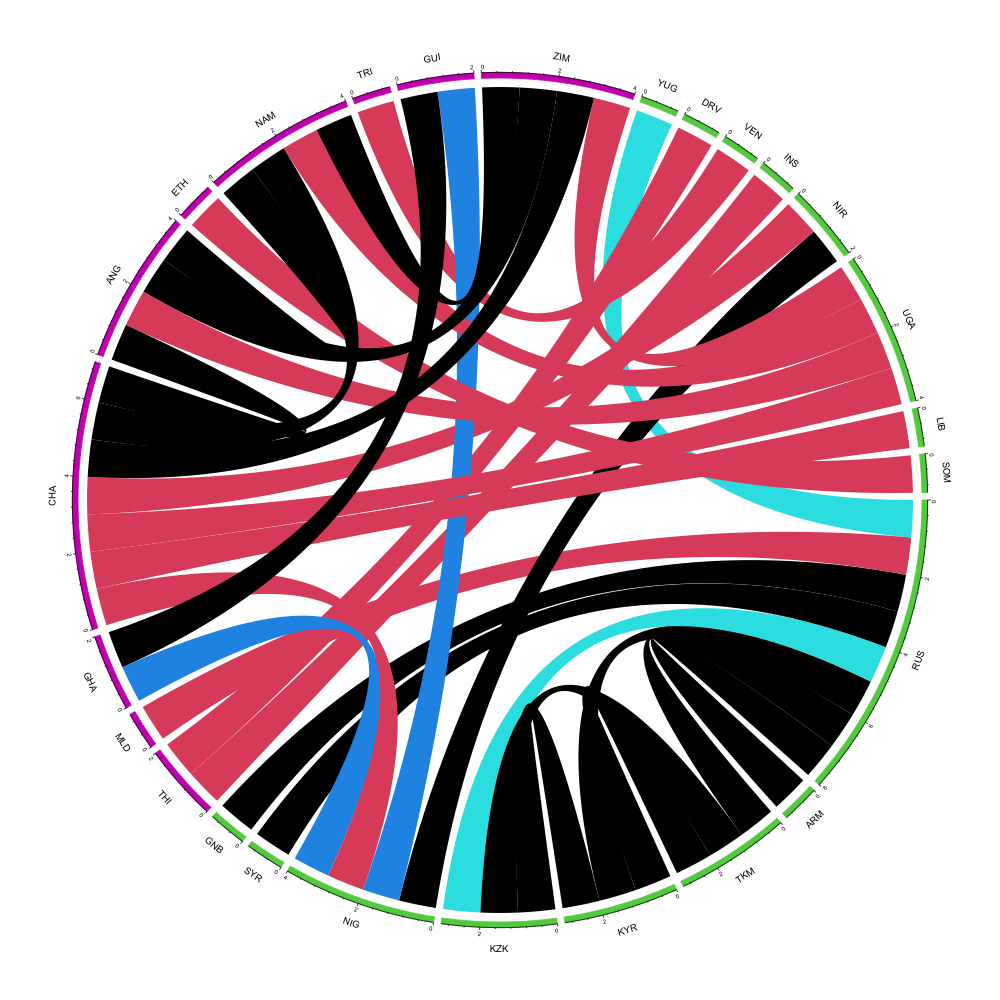}
    \end{subfigure}
 \caption{Chord diagrams between different communities.}
 \label{fig:chord}
\end{figure}

Figure~\ref{fig:chord} gives the chord diagrams of countries in different communities, where black and blue chords represents positive edges within and between communities, and red and green chords denote negative edges between and within communities. Panel (a) of Figure~\ref{fig:chord} contains the 1st (black arcs) and 3rd (green arcs) communities, while panel (b) contains the 3rd (green arcs) and 6th(purple arcs) communities. It is shown that most positive edges occur within communities and most negative edges occur between communities. Besides, there still exist a few anomaly edges, including the negative edges within community (green) and positive edges between communities (blue).

It is also interesting to note that both China and Japan are in the fourth community, but the estimated anomaly effect between them is negative with $\tilde s_{ij} = -0.055$, suggesting a non-constructive relationship between these two countries. Similarly, although Israel and Turkey are both allies of the United States and contained in the first community, the estimated anomaly effect between them is also negative with $\tilde s_{ij} = -0.067$. On positive anomaly, though China and Pakistan or China and Israel are in different communities, the estimated anomaly effect between China and Pakistan is $\tilde s_{ij} = 0.033,$ and that between China and Israel is $\tilde s_{ij} = 0.044$. All these estimated anomaly effects are well expected due to their well-known historical conflicts or friendships.

\section{Discussion}\label{sec:discussion}

In this article, we propose a unified embedding model for signed networks, which is one of the first attempts to incorporate both balance structure and anomaly effect in signed network modeling.
Asymptotic analysis has been conducted to assure estimation consistency of the proposed embedding model. Its applications to community detection and anomaly detection in signed network are also considered, with sound theoretical justification. The advantage of the proposed embedding model is supported by extensive numerical experiments on both synthetic networks and an international relation network. It is also worth noting that the proposed embedding model is flexible and can be extended to various signed networks, such as the directed signed networks or multi-layer signed networks.





\appendix
\section*{Appendix}

\bibliographystyle{apalike}
\bibliography{ref}

\begin{thebibliography}{}

\bibitem[Axelrod and Bennett, 1993]{axelrod1993landscape}
Axelrod, R. and Bennett, D.~S. (1993).
\newblock A landscape theory of aggregation.
\newblock {\em British journal of political science}, 23(2):211--233.

\bibitem[Bansal et~al., 2004]{bansal2004correlation}
Bansal, N., Blum, A., and Chawla, S. (2004).
\newblock Correlation clustering.
\newblock {\em Machine learning}, 56(1):89--113.

\bibitem[Bhaskar, 2016]{bhaskar2016probabilistic}
Bhaskar, S.~A. (2016).
\newblock Probabilistic low-rank matrix completion from quantized measurements.
\newblock {\em The Journal of Machine Learning Research}, 17:2131--2164.

\bibitem[Cand{\`e}s et~al., 2011]{candes2011robust}
Cand{\`e}s, E.~J., Li, X., Ma, Y., and Wright, J. (2011).
\newblock Robust principal component analysis?
\newblock {\em Journal of the ACM (JACM)}, 58(3):1--37.

\bibitem[Cartwright and Gleason, 1966]{cartwright1966number}
Cartwright, D. and Gleason, T.~C. (1966).
\newblock The number of paths and cycles in a digraph.
\newblock {\em Psychometrika}, 31(2):179--199.

\bibitem[Cartwright and Harary, 1956]{cartwright1956structural}
Cartwright, D. and Harary, F. (1956).
\newblock Structural balance: a generalization of heider's theory.
\newblock {\em Psychological review}, 63(5):277.

\bibitem[Chandrasekaran et~al., 2011]{chandrasekaran2011rank}
Chandrasekaran, V., Sanghavi, S., Parrilo, P.~A., and Willsky, A.~S. (2011).
\newblock Rank-sparsity incoherence for matrix decomposition.
\newblock {\em SIAM Journal on Optimization}, 21(2):572--596.

\bibitem[Chatterjee et~al., 2011]{Chatterjee2011rand}
Chatterjee, S., Diaconis, P., and Sly, A. (2011).
\newblock Random graphs with a given degree sequence.
\newblock {\em The Annals of Applied Probability}, 21(4):1400--1435.

\bibitem[Chen et~al., 2014]{chen2014overlapping}
Chen, Y., Wang, X., Yuan, B., and Tang, B. (2014).
\newblock Overlapping community detection in networks with positive and
  negative links.
\newblock {\em Journal of Statistical Mechanics: Theory and Experiment},
  2014(3):P03021.

\bibitem[Chiang et~al., 2014]{chiang2014prediction}
Chiang, K.-Y., Hsieh, C.-J., Natarajan, N., Dhillon, I.~S., and Tewari, A.
  (2014).
\newblock Prediction and clustering in signed networks: a local to global
  perspective.
\newblock {\em The Journal of Machine Learning Research}, 15:1177--1213.

\bibitem[Chiang et~al., 2012]{chiang2012scalable}
Chiang, K.-Y., Whang, J.~J., and Dhillon, I.~S. (2012).
\newblock Scalable clustering of signed networks using balance normalized cut.
\newblock In {\em Proceedings of the 21st ACM international conference on
  Information and knowledge management}, pages 615--624.

\bibitem[Cucuringu et~al., 2019]{cucuringu2019sponge}
Cucuringu, M., Davies, P., Glielmo, A., and Tyagi, H. (2019).
\newblock Sponge: A generalized eigenproblem for clustering signed networks.
\newblock In {\em The 22nd International Conference on Artificial Intelligence
  and Statistics}, pages 1088--1098. PMLR.

\bibitem[Dattorro, 2010]{dattorro2010convex}
Dattorro, J. (2010).
\newblock {\em Convex optimization \& Euclidean distance geometry}.
\newblock Lulu. com.

\bibitem[Davis, 1967]{davis1967clustering}
Davis, J.~A. (1967).
\newblock Clustering and structural balance in graphs.
\newblock {\em Human relations}, 20(2):181--187.

\bibitem[Doreian and Mrvar, 1996]{doreian1996partitioning}
Doreian, P. and Mrvar, A. (1996).
\newblock A partitioning approach to structural balance.
\newblock {\em Social networks}, 18(2):149--168.

\bibitem[Easley et~al., 2010]{easley2010networks}
Easley, D., Kleinberg, J., et~al. (2010).
\newblock {\em Networks, crowds, and markets}, volume~8.
\newblock Cambridge university press Cambridge.

\bibitem[Erd\"os and R\'enyi, 1960]{Erdos1960evol}
Erd\"os, P. and R\'enyi, A. (1960).
\newblock The evolution of random graphs.
\newblock {\em Magyar Tud. Akad. Mat. Kutat\'o Int. K\"ozl}, 5:17--61.

\bibitem[Facchetti et~al., 2011]{facchetti2011computing}
Facchetti, G., Iacono, G., and Altafini, C. (2011).
\newblock Computing global structural balance in large-scale signed social
  networks.
\newblock {\em Proceedings of the National Academy of Sciences},
  108(52):20953--20958.

\bibitem[Graham, 2017]{Graham2017econ}
Graham, B. (2017).
\newblock An econometric model of network formation with degree heterogeneity.
\newblock {\em Econometrica}, 85(4):1033--1063.

\bibitem[Harary, 1953]{harary1953notion}
Harary, F. (1953).
\newblock On the notion of balance of a signed graph.
\newblock {\em Michigan Mathematical Journal}, 2(2):143--146.

\bibitem[Heider, 1946]{heider1946attitudes}
Heider, F. (1946).
\newblock Attitudes and cognitive organization.
\newblock {\em The Journal of psychology}, 21(1):107--112.

\bibitem[Hoff et~al., 2002]{Hoff2002}
Hoff, P.~D., Raftery, A.~E., and Handcock, M.~S. (2002).
\newblock Latent space approaches to social network analysis.
\newblock {\em Journal of the American Statistical Association},
  97(460):1090--1098.

\bibitem[Holland et~al., 1983]{holland1983stochastic}
Holland, P.~W., Laskey, K.~B., and Leinhardt, S. (1983).
\newblock Stochastic blockmodels: First steps.
\newblock {\em Social Networks}, 5:109--137.

\bibitem[Jiang, 2015]{jiang2015stochastic}
Jiang, J.~Q. (2015).
\newblock Stochastic block model and exploratory analysis in signed networks.
\newblock {\em Physical Review E}, 91(6):062805.

\bibitem[Kumar et~al., 2004]{kumar2004simple}
Kumar, A., Sabharwal, Y., and Sen, S. (2004).
\newblock A simple linear time (1+/spl epsiv/)-approximation algorithm for
  k-means clustering in any dimensions.
\newblock In {\em 45th Annual IEEE Symposium on Foundations of Computer
  Science}, pages 454--462. IEEE.

\bibitem[Kunegis et~al., 2009]{kunegis2009slashdot}
Kunegis, J., Lommatzsch, A., and Bauckhage, C. (2009).
\newblock The slashdot zoo: mining a social network with negative edges.
\newblock In {\em Proceedings of the 18th international conference on World
  wide web}, pages 741--750.

\bibitem[Lei and Rinaldo, 2015]{lei2015consistency}
Lei, J. and Rinaldo, A. (2015).
\newblock Consistency of spectral clustering in stochastic block models.
\newblock {\em The Annals of Statistics}, 43:215--237.

\bibitem[Leskovec et~al., 2010]{leskovec2010predicting}
Leskovec, J., Huttenlocher, D., and Kleinberg, J. (2010).
\newblock Predicting positive and negative links in online social networks.
\newblock In {\em Proceedings of the 19th international conference on World
  wide web}, pages 641--650.

\bibitem[Li et~al., 2020]{li2020network}
Li, T., Levina, E., and Zhu, J. (2020).
\newblock Network cross-validation by edge sampling.
\newblock {\em Biometrika}, 107:257--276.

\bibitem[Li et~al., 2014]{li2014comparative}
Li, Y., Liu, J., and Liu, C. (2014).
\newblock A comparative analysis of evolutionary and memetic algorithms for
  community detection from signed social networks.
\newblock {\em Soft Computing}, 18(2):329--348.

\bibitem[Maoz et~al., 2019]{maoz2019dyadic}
Maoz, Z., Johnson, P.~L., Kaplan, J., Ogunkoya, F., and Shreve, A.~P. (2019).
\newblock The dyadic militarized interstate disputes (mids) dataset version
  3.0: Logic, characteristics, and comparisons to alternative datasets.
\newblock {\em Journal of Conflict Resolution}, 63:811--835.

\bibitem[Massa and Avesani, 2005]{massa2005controversial}
Massa, P. and Avesani, P. (2005).
\newblock Controversial users demand local trust metrics: An experimental study
  on epinions. com community.
\newblock In {\em AAAI}, volume~5, pages 121--126.

\bibitem[Moore, 1979]{moore1979structural}
Moore, M. (1979).
\newblock Structural balance and international relations.
\newblock {\em European Journal of Social Psychology}.

\bibitem[Saldana et~al., 2017]{saldana2017many}
Saldana, D.~F., Yu, Y., and Feng, Y. (2017).
\newblock How many communities are there?
\newblock {\em Journal of Computational and Graphical Statistics}, 26:171--181.

\bibitem[Sengupta and Chen, 2018]{sengupta2018block}
Sengupta, S. and Chen, Y. (2018).
\newblock A block model for node popularity in networks with community
  structure.
\newblock {\em Journal of the Royal Statistical Society: Series B (Statistical
  Methodology)}, 80(2):365--386.

\bibitem[Tang et~al., 2016]{tang2016survey}
Tang, J., Chang, Y., Aggarwal, C., and Liu, H. (2016).
\newblock A survey of signed network mining in social media.
\newblock {\em ACM Computing Surveys (CSUR)}, 49(3):1--37.

\bibitem[Traag and Bruggeman, 2009]{traag2009community}
Traag, V.~A. and Bruggeman, J. (2009).
\newblock Community detection in networks with positive and negative links.
\newblock {\em Physical Review E}, 80(3):036115.

\bibitem[Yang et~al., 2007]{yang2007community}
Yang, B., Cheung, W., and Liu, J. (2007).
\newblock Community mining from signed social networks.
\newblock {\em IEEE transactions on knowledge and data engineering},
  19(10):1333--1348.

\bibitem[Zhang et~al., 2021]{zhang2021directed}
Zhang, J., He, X., and Wang, J. (2021).
\newblock Directed community detection with network embedding.
\newblock {\em Journal of the American Statistical Association}, pages 1--11.

\bibitem[Zhao et~al., 2012]{zhao2012consistency}
Zhao, Y., Levina, E., and Zhu, J. (2012).
\newblock Consistency of community detection in networks under degree-corrected
  stochastic block models.
\newblock {\em The Annals of Statistics}, 40(4):2266--2292.

\bibitem[Zheng et~al., 2015]{zheng2015social}
Zheng, X., Zeng, D., and Wang, F.-Y. (2015).
\newblock Social balance in signed networks.
\newblock {\em Information Systems Frontiers}, 17(5):1077--1095.

\end{thebibliography}

\end{document}